\documentclass[epjCONF]{svjour}
\usepackage{graphics}
\usepackage[varg]{txfonts} % Times fonts
\usepackage[latin1]{inputenc}
\session-title{MESON2012 - 12th International Workshop on Meson Production, Properties and Interaction} 
\begin{document}
\title{Test of charge conjugation invariance in the decay of the $\eta$ meson into $\pi^+\pi^-\pi^0$}
\author{M.~Zieli\'{n}ski\inst{1}\fnmsep\thanks{\email{m.zielinski@uj.edu.pl}} \and P.~Moskal\inst{1,2} for the WASA-at-COSY Collaboration}
\institute{Institute of Physics, Jagiellonian University, 30-059 Krak\'{o}w, Poland \and Nuclear Physics Institute, Forschungszentrum J\"{u}lich, D-52425 J\"{u}lich, Germany}
\abstract{
In this work we present the preliminary results of the analysis of the $pp\to pp\eta\to pp\pi^+\pi^-\pi^0$ 
reaction aiming to test the charge conjugation symmetry C in strong interactions. 
Based on approximately $10^5$ identified $\eta\to\pi^+\pi^-\pi^0$ decay events we have extracted
asymmetry parameters sensitive to C symmetry violation for different isospin values of the final state 
and we have established that all are consistent with zero within the obtained accuracy.
} %end of abstract
\maketitle
\section{Introduction}\label{intro}
In the Standard Model of particles and fields, the charge conjugation C along with the spatial parity P
and time reversal T, is one of the most fundamental symmetries. The C operator in quantum field theory 
applied to a particle state $\vert\psi\rangle$, changes all additive quantum numbers of this particle to
opposite sign, leaving the mass, momentum and spin unchanged, and making it an antiparticle state:
\begin{equation}
C\vert\psi\rangle =  \vert\bar{\psi}\rangle.
\end{equation}
In the Quantum Electrodynamics (QED) and Quantum Chromodynamics (QCD) it is postulated 
that C holds in all electromagnetic and strong interactions on the level smaller than $10^{-8}$.
Therefore the C-invariance should imply the balance between the matter and antimatter,
however experimental observations of the Universe shows that there is significantly larger abundance of matter 
over antimatter~\cite{Sakharov:1967dj}. The known CP breaking effect
is insufficient to explain this phenomenon, but it is hoped that investigations of the charge conjugation
invariance may help in clarification of this problem.  

Difficulties in studies of the charge conjugation arise from the fact that
there are only few known particles in nature which are the eigenstates of the C operator. 
The most suitable candidates are neutral and flavorless mesons and the particle-antiparticle systems.
The particularly interesting appears the $\eta$ meson, which plays a crucial 
role for understanding of the low energy Quantum Chromodynamics, and can be also used to tests of the 
fundamental symmetries. 

In the hadronic decay $\eta\to\pi^+\pi^-\pi^0$, the C invariance violation can manifest itself as an 
asymmetry between energy distribution of the $\pi^+$ and $\pi^-$ mesons in the rest frame of the 
$\eta$ meson. The convenient way to study this invariance is to use Dalitz plot described by the 
Mandelstam variables defined as: 
\begin{equation}
s_i = (p_{\eta} - p_{i})^2 = (m_\eta - m_i)^2 - 2\cdot m_\eta T_i,
\end{equation}
where $p_i$ and $m_i$ denote the four-momentum vectors and masses of final state particles, and $T_i$ stands
for the kinetic energy in the rest frame of the $\eta$ meson. For the $\pi^{+}\pi^{-}\pi^{0}$ final state 
where $m_{\pi^{+}} = m_{\pi^{-}}$, one can use the symmetrized and dimensionless variables defined as:
\begin{equation}
X = \sqrt{3}\left(\frac{T_{+} - T_{-}}{Q}\right),~~~~Y = \frac{3T_{0}}{Q} - 1,
\label{dX}
\end{equation}
where $Q = T_{+} + T_{-} + T_{0}$ is the excess energy.
The Dalitz plot distribution inside the kinematic boundaries is symmetric and flat when the 
transition matrix element is constant. However, in general the density distribution is given by the 
matrix element squared which can be described by expanding the amplitude in the powers of X and Y:
\begin{equation}
\vert M \vert^2 = A^2_0 (1 + aY + bY^2 + cX + dX^2 + fY^3 + ...),
\label{amplituda}
\end{equation}
where $a, b, c, d, f, ...$ are the parameters which can be obtained phenomenologically or on the ground of theory, and $A_0$ stands for the normalization factor. 

The amplitude mixing between $\lambda_{C} = -1$ and $\lambda_{C} = +1$, describing the transition 
into isospin state $I=1$ 
and $I=0,2$, respectively, can be investigated by studying of the symmetries of population in different 
parts of the Dalitz plot. In particular the possible presence of C violation could be observed in three
parameters:
(i) left-right asymmetry -- $A_{LR}$, 
(ii) quadrant asymmetry -- $A_{Q}$, and 
(iii) sextant asymmetry -- $A_{S}$. 
Each of these parameters depends on different isospin states of the final three pions. The asymmetries 
are defined as number of events observed in different sectors of the Dalitz plot. 
The left-right asymmetry is defined as: 
\begin{equation}
A_{LR} = \frac{N_{R} - N_{L}}{N_{R} + N_{L}},
\label{ALR}
\end{equation}
where the $N_{L}$ stands for the number of events where $\pi^-$ has a larger energy than 
$\pi^+$ and and $N_{R}$ denotes the number of events where the $\pi^+$ has greater energy than $\pi^-$.
It is sensitive to C violation averaged over all isospin states. However, it is possible to test the 
charge conjugation invariance in given $I$ state. For this, one uses the quadrant and sextant 
asymmetries which are defined as:
\begin{equation}
A_{Q} = \frac{N_{1} + N_{3} - N_{2} - N_{4}}{N_{1} + N_{2} + N_{3} + N_{4}},
\label{AQ}
\end{equation}
\begin{equation}
A_{S} = \frac{N_{1} + N_{3} + N_{5} - N_{2} - N_{4} - N_{6}}{N_{1} + N_{2} + N_{3} + N_{4} + N_{5} + N_{6}},
\label{AS}
\end{equation}
where $N_i$ denotes the number of observed events in $i$-th sector of the Dalitz plot. 
The quadrant asymmetry tests the C invariance in transition into the $3\pi$ final state 
with $I=2$, and the sextant asymmetry is sensitive to the $I=1$~\cite{Jarlskog:2002zz}.  

\section{Experiment}\label{sec:1}
We investigated the $\eta\to\pi^+\pi^-\pi^0$ decay which may violate charge conjugation,
by means of the WASA-at-COSY detector. The $\eta$ meson was produced via $pp\to pp\eta$ reaction
at the proton beam momentum of 2.14~GeV/c. 
The tagging of the $\eta$ meson was done by means of the missing mass technique and the decay products were 
identified by the invariant mass reconstruction.

Two scattered protons were registered in the Forward Detector using the scintillator detectors (FRH and FTH)
and straw tube tracker (FPC), and identified by means of the energy loss method: $\Delta E - E$. 
Charged pions $\pi^+$ and $\pi^-$ were registered in Central Detector using the 
Mini Drift Chamber (MDC) and the four-momenta vectors were reconstructed based on the track curvature in 
the magnetic field of the Superconducting Solenoid. The gamma quanta originating from the $\pi^0$ decay 
were registered in the Scintillating Electromagnetic Calorimeter (SEC). Furthermore, based on the 
reconstruction of the invariant mass of two $\gamma$ quanta, the neutral pion was identified.      

The background originating from the direct two pion production and other $\eta$ meson decays has been reduced to 
negligible level by applying the momentum and energy conservation laws, and by 
using conditions on the missing and invariant mass distributions. The remaining physical background for the 
$\eta\to\pi^+\pi^-\pi^0$ decay originating from the direct production 
of three pions via $pp\to pp\pi^+\pi^-\pi^0$ was subtracted for each studied phase space interval separately.

\section{Results}\label{sec:1}
The asymmetry parameters were determined by dividing the Dalitz plot into regions according to the 
formulas (5), (6) and (7). The events were summed up separately for odd and even regions and a corresponding 
missing mass for the $pp\to pp\eta$ reaction was reconstructed for each region. 
Furthermore, to determine the number of events corresponding to the $\eta\to\pi^+\pi^-\pi^0$ decay 
in each region the background was subtracted using the polynomial fit method, and the correction for 
acceptance and efficiency obtained based on the simulations of signal reaction, was applied.
\begin{figure}[t]
\mbox{
\hspace{-0.3cm}
\resizebox{0.35\columnwidth}{!}{\includegraphics{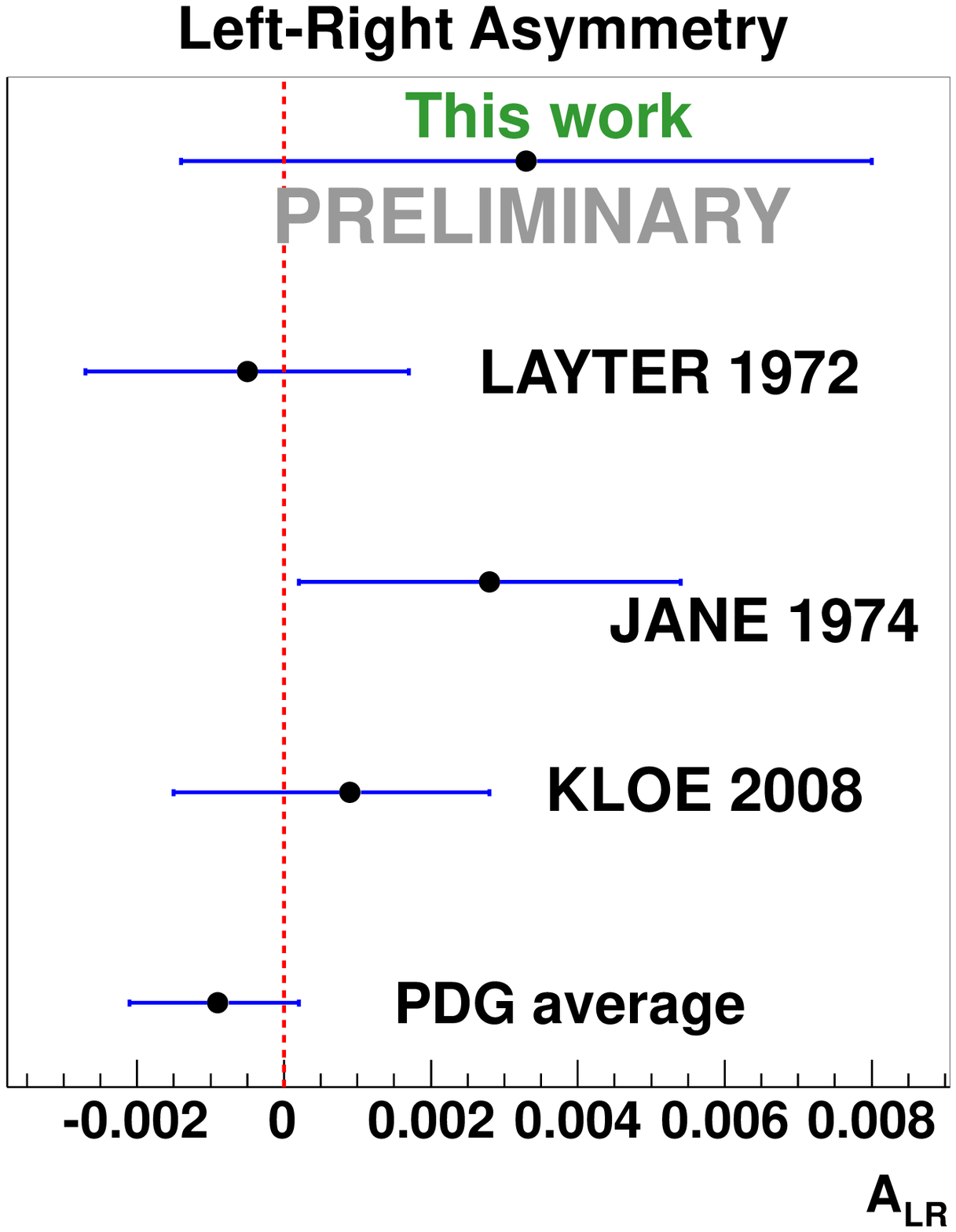}}
\hspace{-0.3cm}
\resizebox{0.35\columnwidth}{!}{\includegraphics{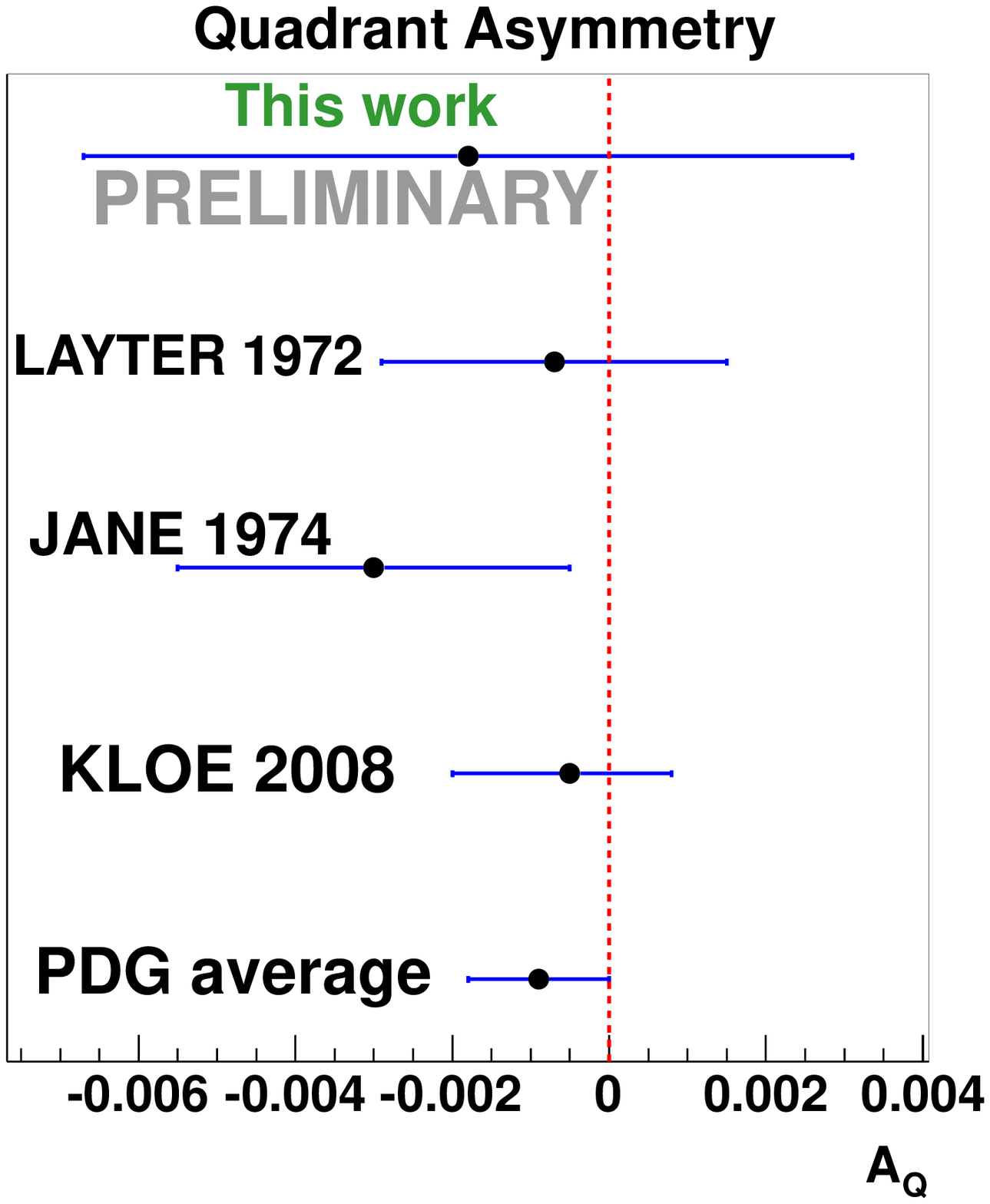}}
\hspace{-0.3cm}
\resizebox{0.35\columnwidth}{!}{\includegraphics{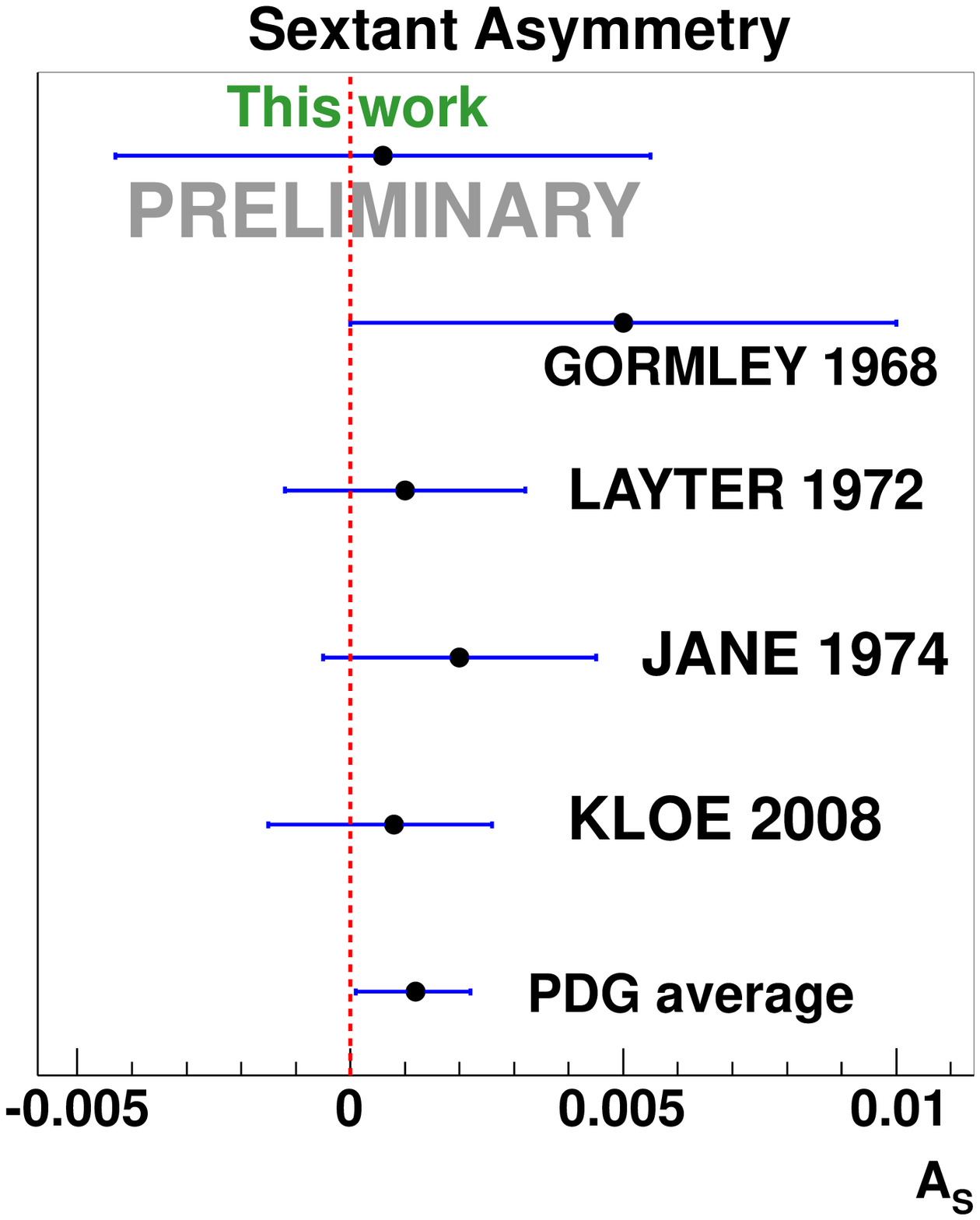}}
}
\caption{Comparison of obtained values of asymmetries~\cite{Zielinski:2012phd} with results determined 
by previous experiments~\cite{Layter:1973ti,Jane:1974mk,Ambrosino:2008ht}, 
and a value given by PDG~\cite{Nakamura:2010zzi}.}
\label{fig:1} 
\end{figure}
The preliminary estimated values of the asymmetries are shown in Fig.~\ref{fig:1}.
Established values of the asymmetry parameters are consistent with
zero within the range of the statistical and systematic uncertainty, which allows to conclude 
that the charge conjugation symmetry C is conserved in strong interactions on the level of the achieved accuracy. Obtained results are also in agreement with previously measured 
values~\cite{Layter:1973ti,Jane:1974mk,Ambrosino:2008ht} and the average of the Particle 
Data Group~\cite{Nakamura:2010zzi} (see Fig.\ref{fig:1}).

\section{Outlook}\label{sec:2}
The WASA-at-COSY currently collected around $10^{9}$ $\eta$ mesons in proton-proton collisions, which is 
one of the world's largest data sample for the $\eta$ meson, therefore the studies on the charge conjugation
invariance in the $pp$ interactions will be continued. Available statistics should enable to lower 
the statistical uncertainties for the determination of the asymmetry parameters by a factor of five in future 
analysis.\\

\noindent
{\bf{Acknowledgments}}\\
This work was supported by the Polish National Science Centre under the Grant Agreement No.\\0312/B/H03/2011/40, 
by the MesonNet, and by the FFE grants from the Forschungszentrum J\"{u}lich.


\begin{thebibliography}{99}
\bibitem{Sakharov:1967dj} A.~D.~Sakharov, Pisma Zh.\ Eksp.\ Teor.\ Fiz.\  {\bf 5} (1967) 32.
\bibitem{Jarlskog:2002zz} C.~Jarlskog and E.~Shabalin, Phys.\ Scripta T {\bf 99} (2002) 23.  
\bibitem{Layter:1973ti} J.~G.~Layter {\it et al.}, Phys.\ Rev.\ D {\bf 7} (1973) 2565. 
\bibitem{Jane:1974mk} M.~R.~Jane {\it et al.}, Phys.\ Lett.\ B {\bf 48} (1974) 260.  
\bibitem{Ambrosino:2008ht} F.~Ambrosino {\it et al.}  [KLOE Collaboration], JHEP {\bf 0805} (2008) 006. 
\bibitem{Nakamura:2010zzi} K.~Nakamura {\it et al.}  [Particle Data Group], J.\ Phys.\ G {\bf 37} (2010) 075021.
\bibitem{Zielinski:2012phd} M.~Zieli\'{n}ski, PhD. Thesis, Jagiellonian University Krak\'{o}w (2012).
\end{thebibliography}
\end{document}